\documentclass[preprint,floatfix] {revtex4} 
\newcommand{\rvec}{\mathrm {\mathbf {r}}} 
\newcommand{\rpvec}{\mathrm {\mathbf {r'}}} 
\usepackage{graphicx}
\usepackage{subfigure}
\begin{document}

\title{A new DFT method for atoms and molecules in Cartesian grid}
\author{Amlan K.\ Roy}
\email{akroy@iiserkol.ac.in, akroy@chem.ucla.edu} 
\affiliation{Division of Chemical Sciences, Indian Institute of Science Education and Research (IISER),
Salt Lake, Kolkata-700106, India}

\begin{abstract}
Electronic structure calculation of atoms and molecules, in the past few decades has largely been 
dominated by density functional methods. This is primarily due to the fact that this can account for 
electron correlation effects in a rigorous, tractable manner keeping the computational cost at a 
manageable level. With recent advances in methodological development, algorithmic progress as well 
as computer technology, larger physical, chemical and biological systems are amenable to quantum mechanical calculations than ever 
before. Here we report the development of a new method for accurate reliable description of atoms, 
molecules within the Hohenberg-Kohn-Sham density functional theory (DFT). In a Cartesian grid,  
atom-centered localized basis set, electron density, molecular orbitals, two-body potentials are directly 
built on the grid. We employ a Fourier convolution method for classical Coulomb potentials by making 
an Ewald-type decomposition technique in terms of short- and long-range interactions. One-body matrix 
elements are obtained from standard recursion algorithms while two-body counterparts are done by direct 
numerical integration. A systematic analysis of our results obtained on various properties, such as 
component energy, total energy, ionization energy, potential energy curve, atomization energy, etc., 
clearly demonstrates that the method is capable of producing quite accurate and competitive (with those 
from other methods in the literature) results. In brief, a new variational DFT method is presented 
for atoms and molecules, \emph{completely} in Cartesian grid. 
\end{abstract}
\maketitle

\section{introduction}
Calculation of wave functions of large molecules by \emph{first principles} methods has been an outstanding 
problem having much relevance in varied fields such as theoretical chemistry, condensed matter physics, 
material science, etc. On the one hand, there are standard Roothaan-Hartree-Fock (RHF)-type methods, as 
implemented in several quantum chemistry program packages, though not difficult are certainly tedious
and cumbersome indeed, if the number of basis functions becomes large (which is easily the case for even
reasonably smaller molecules). Also these methods ignore the important effects arising from electron 
correlation. On the other side, there are numerous semi-empirical methods, which admittedly have often 
found various successful applications in describing molecular properties, but raises many questions regarding
their applicability for some systems such as transition metal complexes with different kind of ligands. They
suffer from the well-known problem of parametrization; stated differently, a certain parametrization scheme
is usually successful for a restricted class of compounds with respect to a restricted number of properties.

Over the last three decades, there has been considerable progress in the formulation and implementation of
density functional methods, of which $X\alpha$ or Hartree-Fock-Slater, is the simplest, best known. The
primary reason for this is because it can account for electron correlation effects in a rigorous, quantitative, 
transparent manner. Moreover it also provides a good compromise between computational cost and accuracy. Some other popular
routes toward introducing electron correlation in a many-electron problem are through Moller-Plesset (MPn) and 
coupled cluster methods. Density functional theory (DFT) \cite{parr89,jones89,dreizler90,chong95,seminario96,
joulbert98, dobson98,nagy98,kohn99,koch01,parr02,fiolhais03,gidopoulos03,martin04,sholl09}, in particular, has 
become a powerful and versatile tool in recent years; and more preferable to the other correlated methods, partly because 
of its favorable 
scaling (which is typically $N^3$, although recently linear-scaling methods have been available). Obviously, the ultimate 
goal is to be able to describe structure, dynamics, properties of larger and larger systems 
as accurately as possible (close to experimental results) with optimal computational resources. Recent dramatic 
explosion in computer technology and emergence of accurate density functional techniques, in both formalism and 
algorithmic aspects, have made it possible to reach this goal which has eluded quantum chemistry for long. 
In more practical terms, this is achieved through a successful marriage of basis-set approaches to electronic 
structure theory and efficient grid-based quadrature schemes to produce a scaling which is at least as good as self-consistent 
methods. This also simultaneously allows the very difficult many-body effects to be approximated by an 
effective \emph{one-electron} potential. A proliferation of DFT-based methodologies has been witnessed for electronic 
structure calculation of a broad range of systems including atoms, molecules, clusters, solids and this continues to 
grow at a rapid pace.

The key concept of DFT is that \emph{all} desired properties of a many-electron interacting system can be obtained in terms
of the ground-state electron density, $\rho(\rvec)$, in stead of a complicated
many-electron wave function, as in traditional \emph{ab-initio} approaches. This real, non-negative, 3D, scalar 
function of position is easily visualizable (in contrast to the wave function which is, in general, complex, 4N
dimensional and not so easily interpretable visually). It has \emph{direct} physical significance (can be directly measured
experimentally) and, in principle, provides all the informations about \emph{ground and all excited states}, as 
obtainable from a wave function. Initial attempts to use electron density as a basic variable for a many-electron
system, is almost as old as quantum mechanics and is due, independently to, Thomas and Fermi \cite{thomas27,fermi27}.
In this quantum statistical model, kinetic energy of an interacting system is approximated as an explicit 
functional of density (by assuming electrons to be in the background of a non-interacting homogeneous electron gas), 
while electron-nuclear attraction and electron repulsion contributions are treated classically. However, since this
completely ignores the important exchange, correlation effects and kinetic energy is approximated very crudely, results
obtained from this method are rather too crude to be of any use. It also fails to explain the essential physics
and chemistry such as shell structure of atoms and molecular binding. Significant improvements were made by 
Dirac \cite{dirac30} by introducing exchange effects into the picture, the so-called local density 
approximation (LDA). 

The stark simplicity of these above procedures encouraged multitude of solid-state and molecular calculations. 
However, due to a lack of rigorous foundation as well as considerably large errors encountered in these works, 
the theory lost its charm and appeal until a breakthrough work by Hohenberg and Kohn \cite{hohenberg64}, which rekindled the hope. 
This changed the status of DFT as it got a firm footing and laid the groundwork of all of today's DFT. The
first theorem simply states that the external potential $v_{ext}(\rvec)$, and hence total energy of an interacting
system is a unique functional of $\rho(\rvec)$. According to the second theorem, ground-state energy can be 
obtained variationally, i.e., the density that minimizes total energy is the exact ground-state density. Note that
although these theorems are very powerful, they merely prove the \emph{existence} of a functional, but do not offer any route
of computing this density in practical terms. Even though a mapping between ground-state density and energy is
established, it remains mute about the construction of this ``universal" functional. Thus as far as computational
DFT is concerned very little progress is made compared to the prevailing situation. One still needs to solve the
many-body problem in presence of $v_{ext}(\rvec)$.  

The situation changed dramatically in following year after the publication of a seminal work by Kohn and Sham
\cite{kohn65}, who proposed a clever route to approach the unknown universal functional. This is done 
by mapping the full interacting system of interest with the real potential onto a fictitious, non-interacting system
of particles. The electrons move in an effective Kohn-Sham (KS) single-particle potential $v_{KS}(\rvec)$ and the
auxiliary system yields same ground-state density as the real interacting system, but greatly simplifies our
calculation. Since exact wave functions of non-interacting fermions are represented by Slater determinants, 
major portion of kinetic energy can be computed to a good accuracy, in terms of one-electron orbitals which
construct the reference system. The residual unknown contribution of kinetic energy (which is a fairly small
quantity) is dumped in to the unknown, non-classical component of electron-electron repulsion as, 
\begin{eqnarray}
F[\rho] & = & T_s[\rho]+J[\rho]+E_{xc}[\rho] \nonumber \\
E_{xc}[\rho] & = & (T[\rho]-T_s[\rho]) + (E_{ee}[\rho]-J[\rho]) = T_c[\rho]+E_{nc}[\rho].
\end{eqnarray}
$T_s[\rho]$ signifies exact kinetic energy of the hypothetical non-interacting system; $J[\rho]$ the classical 
component of electron-electron repulsion. In this partitioning scheme then, $E_{xc}[\rho]$ contains everything 
that is unknown, i.e., non-classical electrostatic effects of electron-electron repulsion as well as the difference
between true kinetic energy $T_c[\rho]$ and $T_s[\rho]$. Now one can write the single-particle KS equation in 
standard form,
\begin{equation}
\left[ -\frac{1}{2} \nabla^2 +v_{eff}(\rvec) \right] \psi_i(\rvec) = \epsilon_i \psi_i(\rvec)
\end{equation}
with the ``effective" potential $v_{eff}(\rvec)$ including following terms, 
\begin{equation}
v_{eff}(\rvec)=v_{ext} (\rvec) + \int \frac{\rho(\rpvec)}{|\rvec-\rpvec|} \ \mathrm{d} \rpvec + v_{xc} (\rvec)
\end{equation}
where $v_{eff}(\rvec)$ and $v_{ext}(\rvec)$ signify the effective and external potentials respectively. Exact 
form of exchange-correlation (XC) functional
remains unknown as yet and its accurate form is necessary for description of real interacting systems (such as 
binding properties). Numerous approximations have been suggested for this and development of improved functionals
has constituted one of the most fertile areas of research for several years and even today. 

For practical purposes, minimization of the explicit functional is not an easy or efficient task; and hence not recommended. 
A far more attractive route is, in stead of working \emph{solely} in terms of density, one might bring back the usual 
orbital picture in to the problem. This gives an appearance like a single-particle theory, albeit incorporating
the many-body effects, in principle, \emph{exactly}. In straightforward \emph{real-space} \cite{laaksonen85,becke89,
white89,tsuchida95,beck97,hernandez97,modine97,kim99,beck00,chelikowsky00,lee00,wang00,kronik06} solution of this equation, 
the wave functions are sampled on a real grid through either of the following three representations such as 
finite difference (FD), finite element (FE) or wavelets where the solution is obtained in an iterative mechanism. 
The advantage is that, in all cases, relevant discrete differential equations offer highly structured, banded 
sparse matrices. Moreover, the potential operator is diagonal in coordinate space; Laplacian operator is nearly local (making 
them ideal candidates for linear-scaling approaches), and these are easily amenable to domain-decomposition parallel 
implementation. One can use adaptive mesh refinements or coordinate transformations to gain further resolution in 
local regions of space. The \emph{whole} molecular grid belongs to either uniform \cite{chelikowsky94a, chelikowsky94b,
chelikowsky00} or refined uniform grids \cite{white89,tsuchida95,beck97,hernandez97,beck00,wang00,kronik06}. The 
classical electrostatic potential can be found using highly optimized FFTs or real-space multigrid algorithms. Using 
FD and FE approach, reasonably successful, fully numerically converged solution for self-consistent KS eigenvalue 
problem of atoms/molecules has been reported in literature \cite{laaksonen85,becke89,white89}. In another development, 
an atom-centered numerical grid \cite{becke89} was proposed for performing molecular-orbital (MO) calculation. The 
physical domain was partitioned into a collection of single-center components with radial grids centered 
at each nucleus. Later, a high-order real-space pseudopotential method \cite{chelikowsky94a,chelikowsky94b} was 
presented for relatively larger systems in uniform Cartesian coordinates. In an orthogonal 3D mesh, an $m$th order FD
expansion of the Laplacian can be written as, 
\begin{equation}
\left[ \frac{\partial^2 \psi}{\partial x^2} \right]_{x_i, y_j, z_k} = \sum_{-m}^{m}
C_m \psi(x_i+mh, y_j, z_k) +O (h^{2m+2}).
\end{equation}
The Hartree potential is obtained by a direct summation on grid by an iterative summation technique. 
FD method has been used in different flavors \cite{chelikowsky94a,chelikowsky94b,modine97,kim99,lee00} for a 
number of interesting \emph{ab initio} self-consistent problems in clusters and other finite systems, such as
forces, molecular dynamics simulations, polarizabilities of semiconductor clusters as well for areas outside
traditional electronic structures. Multigrid methods \cite{briggs95,briggs96,wang00}, which accelerate the 
self-consistent procedure by reducing number of grid points considerably, have found many applications in 
calculations for molecules and large condensed phase systems on uniform grids. Use of these in conjunction with 
adaptive grid to enhance the resolution \cite{modine97,gygi95} has been studied. Convergence is influenced by 
parameters like grid spacing, domain size, order of representation, etc.   

Some of the shortcomings of above approaches are that these are non-variational and dimension of the Hamiltonian 
matrix is unmanageably large. In an alternative approach, finite expansion bases of localized one-electron functions 
is employed, such as exponential functions (STO), Gaussian type orbitals (GTO), plane waves (PW), wavelets, 
numerical basis sets, linear muffin-tin orbitals, delta functions or some suitable combinations of these. With
STOs or other numerical orbitals, relevant multicenter integrals in the Hamiltonian needs to be evaluated 
numerically, while with a Gaussian basis, these and all other integrals required to compute the matrix elements 
of Hamiltonian, can be obtained analytically. One
pays a price for using the latter; for considerably larger number of such functions are required for accurate
description of electronic states, as they do not exhibit correct behavior at either small or large distances
from nuclei. Gaussian bases have been extensively used in quantum chemistry calculations of small and medium
molecules, whereas PWs (frequently coupled with pseudopotentials to treat core electrons) have been most successful 
for solids. PWs share with Gaussians the same property that the integrals are known analytically. However, unlike 
Gaussians, Coulomb interaction is local in Fourier space---hence solving Poisson's equation, a very important step 
in any DFT calculation, is quite trivial in a PW basis. The two notable disadvantages of PW bases are: (i) periodic
boundary conditions must be used, which is desirable for solids, but not so in case of clusters or molecules (ii)
resolution of the basis is exactly same everywhere. Thus for atoms and molecules Gaussian bases have been most
successful and popular. Combination of GTO and PW bases have also been used \cite{lippert99,krack00}, whereby KS MOs are
expanded in Gaussian bases and the electron density in an augmented PW basis. 

Most of the modern DFT programs, for routine calculations, typically employ the so-called atom-centered 
grid (ACG), pioneered by Becke \cite{becke88}, where a molecular grid is efficiently described in terms of some 
suitable 3D numerical quadratures. This is not necessarily the best strategy, but is a relatively simpler well-defined
path, adopted by majority of DFT programs. The basic step consists of partitioning a molecular integral into
single-center discrete overlapping atomic components. For an arbitrary integrand $F(\rvec)$, such a decomposition 
provides the value of integral $I$ as,  
\begin{equation}
I=\int F(\rvec) \mathrm{d} \rvec = \sum_A^M I_A, \ \ \ \ \ \ \ \mathrm{for} \ M \ \mathrm{nuclei}
\end{equation} 
such that the atomic integrand $F_A$, when summed over all nuclei, returns our original function. Single-center 
atomic contributions are denoted by $I_A= \int F_A(\rvec) \ d \rvec$.
\begin{equation}
\sum_A^M F_A (\rvec) = F(\rvec). 
\end{equation}
$F_A (\rvec)$s are typically constructed from original integrand by some well-behaved weight functions $F_A(\rvec) =
w_A F(\rvec)$. 
The atomic grid constitutes of 
a tensor product between radial part defined in terms of some quadrature formulas such as Gauss-Chebyshev, Gaussian, 
Euler-McLaurin, multi-exponential numerical, etc., \cite{gill93,murray93,treutler95,mura96,krack98,lindh01,gill03,
chien06} and Lebedev angular quadratures (order as high as 131 has been reported, although usually much lower orders 
suffice; 59th order being the one most frequently used) \cite{lebedev75,lebedev76,lebedev92,lebedev94,lebedev1999}. 
Once $F_A$s are determined, $I_A$s are computed on grid as follows (in polar coordinates), 
\begin{equation}
I_A= \int_{0}^{\infty} \int_{0}^{\pi} \int_{0}^{2\pi} F_A(r,\theta,\phi) r^2 \sin \theta 
\ \mathrm{d} r \ \mathrm{d}\theta \ \mathrm{d}\phi \approx \sum_{p}^{P} w_p^{\mathrm{rad}}
\sum_{q}^{Q} w_q^{\mathrm{ang}} \ F_A(r_p, \theta_q, \phi_q),
\end{equation}
where $w_p^{\mathrm{rad}}$, $w_q^{\mathrm{ang}}$ signify radial, angular weights respectively with $P$, $Q$ points 
(total number of points being $P \times Q$). Usually angular part is not further split into separate $\theta$, 
$\phi$ contributions as surface integrations on a sphere can be done numerically quite easily accurately by the help 
of available highly efficient algorithms. Also angular integration has been found to be much improved by Lobatto 
scheme \cite{treutler95}. Many 
variants of this integration scheme have been proposed thereafter, mainly to prune away any extraneous grid points,
which is much desirable and useful. Integration by dividing whole space and invoking product Gauss rule 
\cite{boerrigter88} has been suggested as well. A variational integration scheme \cite{pederson90} divides molecular 
space into three different regions such as atomic spheres, excluded cubic region and interstitial parallelepiped. In 
a Fourier transform Coulomb and multi-resolution technique, both Cartesian coordinate grid (CCG) and ACG were
used \cite{molnar05,brown06,kong06}; former divides Gaussian shell pairs into ``smooth" and ``sharp" categories on the 
basis of exponents while latter connects these two by means of a divided-difference polynomial interpolation to 
translate density and gradients from latter to former. Among other methods, a partitioning scheme \cite{ishikawa99},
linear scaling \cite{stratmann96} and adaptive integration schemes \cite{yamamoto97,krack98,ishikawa99} are 
worth mentioning. 

The purpose of this article is to present an alternate DFT method for atoms and molecules by using a linear 
combination of GTO expansion for the KS molecular orbitals within CCG \emph{solely} \cite{roy08,roy08a,roy08b}, 
that has been developed by this author during the past three years. No auxiliary basis 
set is invoked for charge density. Quantities such as localized atom-centered basis functions, MOs, electron density 
as well as classical Hartree and non-classical XC potentials are constructed on the 3D real grid directly. A Fourier 
convolution method, involving a combination of FFT and inverse FFT \cite{martyna99,minary02} is used to obtain the
Coulomb potential quite accurately and efficiently. Analytical one-electron Hay-Wadt-type effective core potentials 
\cite{wadt85,hay85}, which are made of sum of Gaussian type functions, are used to represent the inner core electrons 
whereas energy-optimized truncated Gaussian bases are used for valence electrons. The validity and performance of our
method is demonstrated in detail for a modest number of atoms and molecules by presenting total energy, energy components, orbital
energy, potential energy curve, atomization energy for both local \cite{vosko80} and non-local Becke exchange 
\cite{becke88a}+Lee-Yang-Parr (LYP) \cite{lee88} correlation energy functionals. Success of these above mentioned
functionals for various physical, chemical processes are well known. However, in absence of the exact functional form although 
these provide very good estimates, in many occasions they behave rather poorly. Thus construction of more accurate elaborate  
sophisticated XC functional lies at the forefront of current research activity as evidenced by an enormous amount of
literature on this. Some frequently used functionals in recent years are generalized gradient expansion, hybrid, 
meta or orbital-dependent functionals, etc. (see, for example, \cite{cramer04}, for a brief review). In order to expand
the scope, applicability and feasibility of our method, we have employed two other relatively lesser used functionals
holding good promise, \emph{viz.}, 
Filatov-Thiel (termed as FT97 in the community) \cite{filatov97,filatov97a} and PBE \cite{perdew96} functionals, which
have been used in many applications with quite decent success. 
Detailed comparisons are made with the widely used GAMESS quantum chemistry
program \cite{schmidt93} (grid-based method in ACG and gridless method), and wherever possible, with experimental results as well. 
While this work exclusively deals with pseudopotential studies, full calculations may be investigated in future communications.
The article is organized as follows. Section II gives a brief summary of the methodology. A discussion on our results is 
presented in Section IV, while we end with a few concluding remarks in Section V. 

\section{Methodology and computational considerations}
Details of this method has been published elsewhere \cite{roy08,roy08a,roy08b}; here we summarize only the 
essential steps. Our starting point is the single-point KS equation for a many-electron system, which, under the 
influence of pseudopotentials, can be written as (henceforth atomic units employed unless otherwise mentioned), 
\begin{equation}
\left[ -\frac{1}{2}\nabla^2 +v^{p}_{ion}(\mathbf{r})+v_H[\rho](\mathbf{r}) +v_{XC}[\rho](\mathbf{r})
\right]
\psi_i(\mathbf{r}) = \epsilon_i \psi_i(\mathbf{r}).
\end{equation}
Here $v^p_{ion}$ denotes the ionic pseudo-potential for the system as,
\begin{equation}
v^{p}_{ion}(\mathbf{r})=\sum_{\mathbf{R_a}} v^{p}_{ion,a}(\mathbf{r}-\mathbf{R_a})
\end{equation}
with $v^{p}_{ion,a}$ signifying the ion-core pseudopotential associated with an atom $A$, situated at
$\mathbf{R_a}$. $v_H[\rho](\mathbf{r})$ describes the classical Hartree electrostatic interactions among valence electrons, 
while $v_{XC}[\rho(\rvec)]$ represents the non-classical XC part of the Hamiltonian, 
which normally depends on electron density (and also probably gradient and other derivatives), but
not on wave functions explicitly. $\{\psi_i^{\sigma}(\mathbf{r})\}, \sigma=\alpha\  \textrm{or}\ \beta,$ corresponds 
to the set of N occupied orthonormal MOs, to be determined from the solution of this equation.


As already hinted, the so-called, linear combination of atomic orbitals (LCAO) ansatz is by far, the most popular, 
convenient and practical route towards an iterative solution of molecular KS equation. In this scheme, the unknown KS MOs 
$\{\psi_i^{\sigma}(\rvec)\}, \sigma=\alpha,\beta$ are linearly expanded in terms of a set of K known basis functions as,
\begin{equation}
\psi_i^{\sigma}(\rvec) = \sum_{\mu=1}^K C_{\mu i}^{\sigma} \chi_{\mu} (\rvec), \ \ \ i=1,2,\cdots,K, 
\end{equation}
where set $\{ \chi_{\mu}(\rvec)\}$ denotes the contracted Gaussian functions centered on constituent atoms while 
$\{C_{\mu i}^{\sigma}\}$ contains contraction coefficients for the orbital $\psi_i^{\sigma}(\rvec)$. The above
expression is exact for a complete set $\{\chi_{\mu}\}$ with $K=\infty$ and, in principle, any complete set
could be chosen. However for practical purposes, infinite basis set is not feasible and one is restricted to a finite
set; thus it is of utmost importance to choose suitable basis functions such that the approximate expansion reproduces
unknown KS MOs as accurately as possible. The procedure is very similar to that applied in HF theory and more 
practical details could be found in the elegant books \cite{szabo96,helgaker00,jensen07}. Individual spin-densities 
are then given by,
\begin{equation}
\rho^{\sigma}(\rvec) = \sum_i^{N^{\sigma}} |\psi_i^{\sigma}(\rvec)|^2 = 
\sum_i^{N^{\sigma}} \sum_{\mu =1}^K \sum_{\nu =1}^K C_{\mu i}^{\sigma} C_{\nu i}^{\sigma} \chi_{\mu} (\rvec) 
\chi_{\nu}^* (\rvec) =  
\sum_{\mu} \sum_{\nu} P_{\mu \nu}^{\sigma} \chi_{\mu} (\rvec) \chi_{\nu}^* (\rvec), 
\end{equation}
where $P^{\sigma}$ stands for the respective density matrices. Denoting the one-electron KS operator in parentheses 
of Eq.~(5) by $\hat{f}^{KS}$, one can write the KS equation in following operator form, 
\begin{equation}
\hat{f}^{KS} \psi_i(\rvec) = \epsilon_i \psi_i (\rvec).
\end{equation}
This operator differs from another similar Fock operator $\hat{f}^{HF}$, used in HF theory, in the sense 
that former includes all non-classical many-body effects arising from electron-electron interaction through XC 
term (as a functional derivative with respect to density, $v_{xc}[\rho] = \delta E_{xc}[\rho] / \delta \rho $), 
whereas there is no provision for such effects in the latter. This represents a fairly complicated system of coupled
integro-differential equation whose numerical solution is far more demanding and some details are mentioned in the following.  

In a spin-unrestricted formalism, substitution of energy terms in to the energy expression, followed by a minimization
with respect to unknown coefficients $C_{\mu i}^{\sigma}$, with 
$\rho(\rvec)= \rho^{\alpha} (\rvec)+ \rho^{\beta} (\rvec)$ and $P=P^{\alpha}+P^{\beta}$, leads to the following matrix 
KS equation, which is reminiscent of Pople-Nesbet equation in HF theory,
\begin{equation}
F^{\alpha} C^{\alpha} =S C^{\alpha} \epsilon^{\alpha},  \ \ \ \ \ \ \ \mathrm{and} \ \ \ \ \ \ \ 
F^{\beta} C^{\beta} =S C^{\beta} \epsilon^{\beta},
\end{equation}
with the orthonormality conditions,
\begin{equation}
(C^{\alpha})^{\dagger} S C^{\alpha} = I, \ \ \ \ \ \ \  \mathrm{and} \ \ \ \ \ \ \ 
(C^{\beta})^{\dagger} S C^{\beta} = I.
\end{equation}
Here $C^{\alpha}$, $C^{\beta}$ are matrices containing MO coefficients, S is the atomic overlap matrix, and 
$\epsilon^{\alpha}, \epsilon^{\beta}$ are diagonal matrices of orbital eigenvalues.
$F^{\alpha}$, $F^{\beta}$ are KS matrices corresponding to $\alpha, \beta$ spins respectively, having matrix 
elements as,
\begin{equation}
F_{\mu \nu}^{\alpha}= \frac{\partial E_{KS}}{\partial P_{\mu \nu}^{\alpha}} =H_{\mu \nu}^{\mathrm{core}} 
+J_{\mu \nu}+ F_{\mu \nu}^{XC\alpha}, \ \ \ \ \mathrm{and} \ \ \ \ 
F_{\mu \nu}^{\beta}= \frac{\partial E_{KS}}{\partial P_{\mu \nu}^{\beta}} =H_{\mu \nu}^{\mathrm{core}} 
+J_{\mu \nu}+ F_{\mu \nu}^{XC\beta}.
\end{equation}
Here $H_{\mu \nu}^{\mathrm{core}}$ represents the bare-nucleus Hamiltonian matrix accounting for one-electron
energies including contributions from kinetic energy plus nuclear-electron attraction. $J_{\mu \nu}$
denotes matrix elements from classical Coulomb repulsion whereas the third term signifies same for non-classical 
XC effects. Obviously, this last one constitutes the most difficult and challenging part of whole SCF process. 

At this stage, it is noteworthy that basis-set HF method scales as $N^4$ (total number of two-electron integrals
with N basis functions), while KS calculations do so no worse than $N^3$. There have been attempts to develop $N^2$
or $N \log N$ scaling algorithms by taking into effect the negligible overlap among basis functions involved.
In some earlier LCAO-MO-based KS DFT implementations in GTO bases \cite{andzelm92}, an \emph{auxiliary} basis 
set (in addition to the one used for MO expansion) was introduced to fit (by some least square or other technique) 
some computationally intensive terms to reduce the integral overhead, making it an $N^3$ process. In one such 
development \cite{sambe75,dunlap79,dunlap79a}, the electron density and XC potential were expanded in terms of two
auxiliary bases $f_i$, $g_j$ respectively as, 
\begin{eqnarray}
\rho(\rvec) & \approx & \tilde{\rho}(\rvec)   =  \sum_i a_i f_i(\rvec) \\
v_{xc}(\rvec) & \approx & \tilde{v}_{xc}(\rvec)   =  \sum_j b_j g_j(\rvec). \nonumber
\end{eqnarray}
Here the fitted quantities are identified with tildes while $\{a_i\}$, $\{b_j\}$, the fitting coefficients, are 
determined by minimization of either a straightforward function of following form, 
\begin{equation}
Z=\int [\rho(\rvec)-\tilde{\rho}(\rvec)]^2 \mathrm{d}\rvec,
\end{equation}
or Coulomb self-repulsion of residual density. Both are subject to the constraint that normalization of fitted density
gives total number of electrons. Originally this technique was first suggested in the context of STOs \cite{baerends73} 
and later extended to GTOs \cite{sambe75}. Matrix elements of XC potential (calculated in real-space) were evaluated by
some suitable analytical means.   

Although this route gained some momentum and was quite successful for many applications, it suffers from some noteworthy
difficulties: (i) many distinct fitting techniques with varied flavors (variational or non-variational) produce some 
inconsistency among various implementations (ii) density and XC fitting constraints differ from method to method (iii) 
fitting density does not automatically preserve the conservation of total number of electrons (iv) such an approach 
considerably complicates the analytic derivative theories. However, at the outset, it may be noted that the main 
reason for such schemes was primarily due to a lack of efficient method for good-quality multi-center integrals. However, 
the last few decades has seen emergence of a huge number of elegant efficient high-quality quadrature schemes for 
such integrals offering very accurate results (see, for example, \cite{gill94}, for a lucid review). 

Unlike the exchange integrals in HF theory (which are analytically evaluated within a GTO basis), KS theory involves 
far more challenging non-trivial integrals (due to their complicated algebraic forms). These are not amenable to direct
analytic route and resort must be taken to numerical methods. 

In this work, the basis functions and MOs are directly built on a real, uniform 3D Cartesian grid simulating a cubic 
box as,
\begin{equation}
r_i =r_0 + (i-1)h_r, \ \ \ \ i =1,2, \cdots , N_r; \ \ \ \ \ \mathrm{for} \ \  r \in \{x,y,z\},
\end{equation}
where $h_r, N_r$ denote the grid spacing and number of grid points respectively ($r_0=-N_r h_r/2$).
The classical electrostatic repulsion as well as XC potentials need to be computed on the real grid. For finite 
systems, possibly the simplest and crudest way to compute $v_H(\rvec)$ is by direct numerical integration. This, in 
general, does not perform efficiently and is feasible only for relatively smaller systems. The preferred option is to
solve the corresponding Poisson equation. An alternate accurate technique, found to be quite successful in molecular
modeling and used here, involves conventional Fourier convolution method and some variants \cite{martyna99,minary02}, 
\begin{eqnarray}
\rho(\mathrm{\mathbf{k}}) & = & \mathrm{FFT} \{ \rho(\rvec) \}  \\
v_H(\rvec) & = & \mathrm{FFT}^{-1}\{ v_H^c(\mathrm{\mathbf{k}}) \rho(\mathrm{\mathbf{k}}). \nonumber
\end{eqnarray}
Here $\rho(\mathrm{\mathbf{k}})$ and $v_H^c(\mathrm{\mathbf{k}})$ represent Fourier integrals of
density and Coulomb interaction kernel respectively in the grid. The former is obtained 
from a discrete Fourier transform of its real-space value by standard FFT quite easily. Evaluation of 
the latter, however, is a non-trivial task because of the presence of singularity in real space and 
demands caution. This is overcome by applying a decomposition of the kernel into long- and 
short-range interactions, reminiscent of the commonly used Ewald summation technique in condensed
matter physics, 
\begin{equation}
v_H^c(\rvec) = \frac{\mathrm{erf}(\alpha r)} {r} + \frac{\mathrm{erfc} (\alpha r)}{r} \equiv
v^c_{H_{\mathrm{long}}} (\rvec) + v^c_{H_{\mathrm{short}}} (\rvec),
\end{equation}
where erf(x) and erfc(x) correspond to error function and its complement respectively. Short-range 
Fourier integral can be calculated analytically; the long-range contribution can be obtained 
directly from FFT of real-space values. There are several other routes as well available for classical 
repulsion as needed in the large-scale electronic structure within KS DFT framework. More thorough account on this 
topic can be found in the review \cite{beck00}. 

All one-electron contributions of Fock matrix including overlap, kinetic-energy, nuclear-electron attraction as well 
pseudopotential matrix elements are completely identical to those encountered in HF calculation; these are obtained by
standard recursion algorithms \cite{obara86,mcmurchie78,kahn76}. Corresponding two-electron matrix elements in real-grid 
are computed through direct numerical integration in the CCG,
\begin{equation}
\langle \phi_{\mu} (\rvec) | v_{HXC} (\rvec) | \phi_{\nu} (\rvec) \rangle = h_x h_y h_z 
\sum_{\mathrm{grid}} \phi_{\mu} (\rvec) v_{HXC} (\rvec) \phi_{\nu} (\rvec).
\end{equation}
The matrix eigenvalue problem is accurately and efficiently solved using standard LAPACK routines \cite{anderson99} 
following usual self-consistent procedure iteratively. The KS eigenfunctions and eigenvalues then give total energies 
and/or other quantities in the standard manner. Convergence of the solution was monitored through (i) potential (ii) total 
energies and (iii) eigenvalues. Tolerance of $10^{-6}$ a.u., was employed for (ii), (iii), while $10^{-5}$ a.u., for (i).

\section{Results and Discussion}
Table I, at first, displays various quantities for a representative molecule Cl$_2$, in its ground state, at an internuclear 
distance of 4.20 a.u. Non-relativistic energies as well as other components and total integrated electron density N are 
given, for LDA XC potential for different CCG sets, as indicated by grid spacing, h$_r$ and number of grid points, N$_r$ 
($r \in x,y,z$). Several combinations of grid parameters were tested and finally 8 of them are presented here, which is 
sufficient to illustrate the general important features. Most of these quantities (barring $E_h, E_x, E_c$) are directly 
comparable with the commonly used versatile GAMESS computer program \cite{schmidt93} using same basis function, XC potential and 
effective core potential. All the LDA calculations referred in this work correspond to the homogeneous electron-gas correlation 
of Vosko-Wilk-Nusair (VWN) \cite{vosko80}. The Hay-Wadt valence basis set employed here, splits valence orbital into inner 
and outer components described respectively by two and one primitive Gaussians. Reference values are obtained from two 
different options, \emph{viz.}, ``grid" and ``grid-free" DFT. Former uses the default ``army" grade with Euler-McLaurin 
quadratures for radial integration and Gauss-Legendre quadrature for angular integration, while the latter \cite{zheng93,
glaesemann98} works through a resolution of identity to facilitate evaluation of relevant molecular integration over functionals
rather than quadrature grids. As the name implies there is no ``grid" in the latter and in a sense, it is quite attractive, as 
there is no complication that arises from finite grid and associated error. However, there is a price to pay in the form
of an \emph{auxiliary} basis set to expand the identity which itself suffers from the same completeness problem. This table 
illustrates many important points which are explained in detail in \cite{roy08}. Here we mention some of the most significant 
observations. First note that Set A shows maximum deviation from reference values for all quantities presumably because 
the box is not large enough to account for all interactions
present in the system. Set B, encompassing a larger box, expectedly offers better results than those in Set A. Interestingly, 
Sets B,C,F produce very similar results for all the quantities, as they all cover same dimensions; however, results for the last
two sets match more closely with each other and differ from Set B, which probably suffers (as reflected in component energies 
as well as N) due to the crudeness of a coarser grid. Sets D,E both produce very good agreement with reference results for all 
the quantities. 
In order to test convergence, some extra calculations are done in a much extended grids G,H, which of course, produces very 
little change. The above discussion leads us
to conclude that D,E,G,H are our four best results, while the first two are sufficiently accurate for all practical 
purposes. For more of these on Cl$_2$ and similar discussion on HCl, consult \cite{roy08}.   

\begingroup
\squeezetable
\begin{table}      
\caption{\label{tab:table1} Comparison of various energy components of Cl$_2$ with reference values at $R=4.20$. CCG and ACG results 
are given in a.u.}
\begin{ruledtabular}
\begin{tabular} {lccccccccc}
    & \multicolumn{2}{c}{$N_r=32$}  &  \multicolumn{3}{c}{$N_r=64$}   & \multicolumn{2}{c}{$N_r=128$}  &  
\multicolumn{1}{c}{$N_r=256$}   & Ref. \cite{schmidt93}    \\
\cline{2-3} \cline{4-6} \cline{7-8} \cline{9-9} 
Set  & A & B & C & D &   E        &  F         &    G       &  H   &   \\
$h_r$                      &  0.3       &   0.4      &   0.2      & 0.3        & 0.4        & 0.1        
                           & 0.2        & 0.1        &            \\
$\langle T \rangle$        &  11.00750  & 11.17919   &  11.18733  & 11.07195   & 11.06448   & 11.18701   
                           & 11.07244   & 11.07244   & 11.07320    \\
$\langle V^{ne}_t \rangle$ & $-$83.43381& $-$83.68501& $-$83.70054& $-$83.45722& $-$83.44290& $-$83.69988
                           & $-$83.45810& $-$83.45810& $-$83.45964 \\
$\langle E_h \rangle$      & 37.94086   & 36.82427   &  36.83193  & 36.58714   & 36.57918   & 36.83133   
                           & 36.58747   & 36.58747   &        \\
$\langle E_x \rangle$      & $-$4.86173 & $-$4.86641 & $-$4.86778 & $-$4.84360 & $-$4.84245 & $-$4.86771 
                           & $-$4.84374 & $-$4.84373 &        \\
$\langle E_c \rangle$      & $-$0.73575 & $-$0.73521 & $-$0.73530 & $-$0.73374 & $-$0.73366 & $-$0.73530 
                           & $-$0.73374 & $-$0.73374 &            \\
$\langle V^{ee}_t \rangle$ & 32.34338   & 31.22265   & 31.22885   & 31.00981   & 31.00306   & 31.22832   
                           & 31.01000   & 31.01000   &  31.01078   \\
$\langle E_{nu} \rangle$   &  11.66667  &  11.66667  & 11.66667   & 11.66667   & 11.66667   & 11.66667   
                           & 11.66667   & 11.66667   &  11.66667 \\
$\langle V \rangle$        & $-$39.42376& $-$40.79570& $-$40.80503& $-$40.78074& $-$40.77317& $-$40.80489
                           & $-$40.78144& $-$40.78144&  $-$40.78219\\
$\langle E_{el} \rangle$   & $-$40.08293& $-$41.28318& $-$41.28437& $-$41.37545& $-$41.37535& $-$41.28455
                           & $-$41.37566& $-$41.37566&  $-$41.37566\\
$\langle E \rangle$        & $-$28.41626& $-$29.61651& $-$29.61770& $-$29.70878& $-$29.70868& $-$29.61789
                           & $-$29.70900& $-$29.70900&  $-$29.70899\footnotemark[1] \\
 $N$                       & 13.89834   & 13.99939   &  13.99865  & 14.00002   &  14.00003  & 13.99864   
                           &  14.00000     &  13.99999         & 13.99998  \\
\end{tabular}
\end{ruledtabular}
\footnotetext[1]{This is from grid-DFT calculation; corresponding grid-free DFT value is 
$-$29.71530 a.u.} 
\end{table}
\endgroup

At this stage, a similar kind of comparison is made for eigenvalues of Cl$_2$ and HCl in various grids keeping same basis 
function, LDA XC potential and pseudopotentials \cite{roy08}. These are rather less sensitive compared
to the quantities discussed above. For all sets, these either match completely with literature values or show a maximum deviation 
of only 
0.0001-0.0007 a.u. Next, potential energy curves for Cl$_2$ and HCl are shown for several sets along with literature results in 
Fig.~I. Total energies are calculated at following ranges of internuclear separation; for Cl$_2$, $R=3.50-5.00$ and for HCl,
$R=1.60-3.10$ a.u., both sampled at intervals of 0.10 a.u. It is gratifying that in both occasions CCG results show very close agreement 
with reference values for entire region of $R$. Set D energies for Cl$_2$ are quite well (higher from the literature 
by only 0.0001 a.u.) up to $R=4.00$ a.u., and thereafter shows a tendency to deviate gradually (the maximum discrepancy being
quite small though; 0.0007 a.u. for $R=5.00$ a.u.). Sets G,I for Cl$_2$ either match completely with reference values or deviates by
a maximum of 0.0001 a.u. Furthermore, we find that our computed energies are always above the reference values, except in
two occasions ($R=4.00$ and 4.30 for Set G). In HCl, all the sets produce very nice agreement with literature results with
Set D performing best. For a more complete discussion see \cite{roy08}. 

\begin{figure}
\begin{minipage}[c]{0.40\textwidth}
\centering
\includegraphics[scale=0.45]{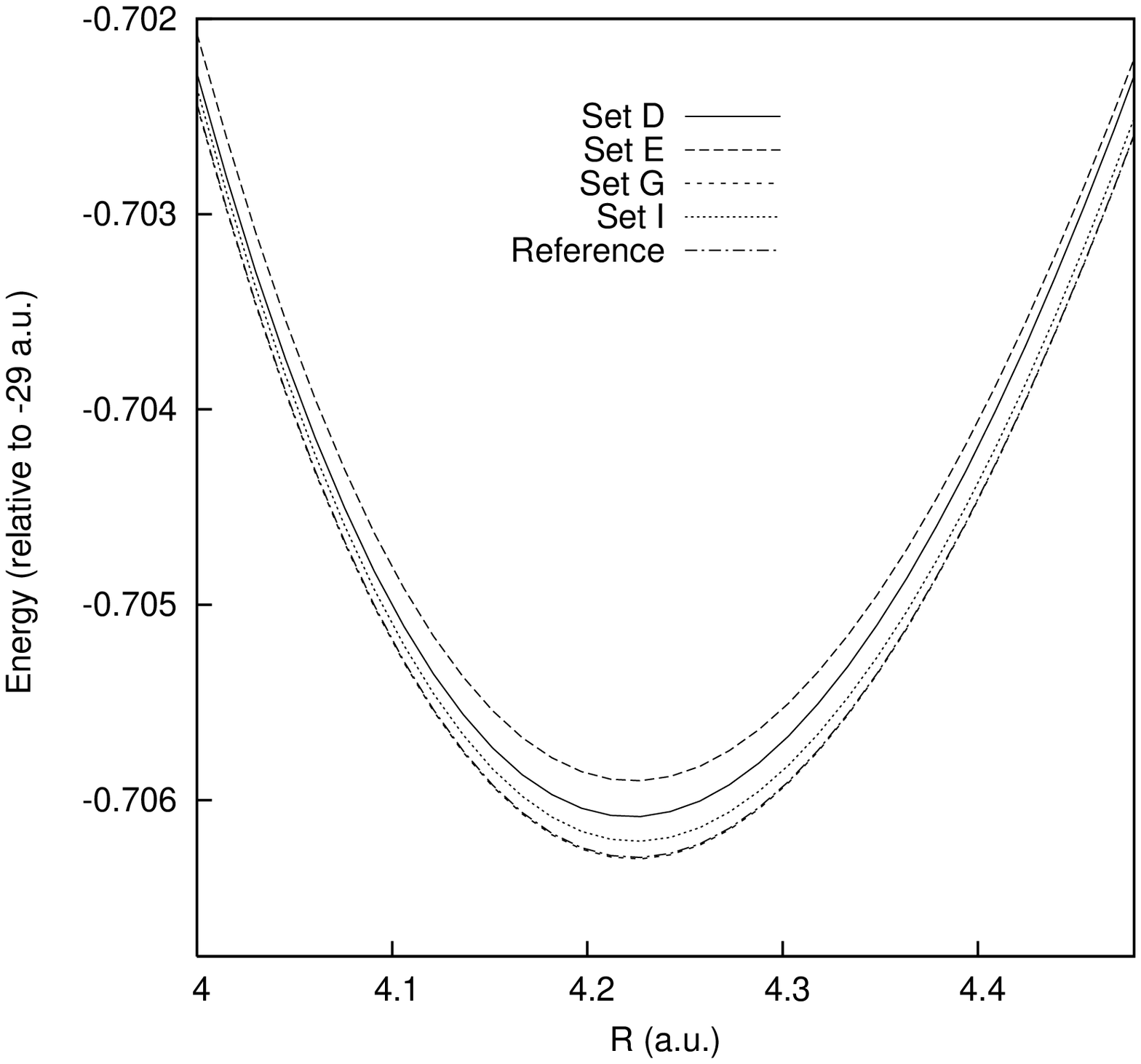}
\end{minipage}%
\hspace{0.5in}
\begin{minipage}[c]{0.40\textwidth}
\centering
 \includegraphics[scale=0.45]{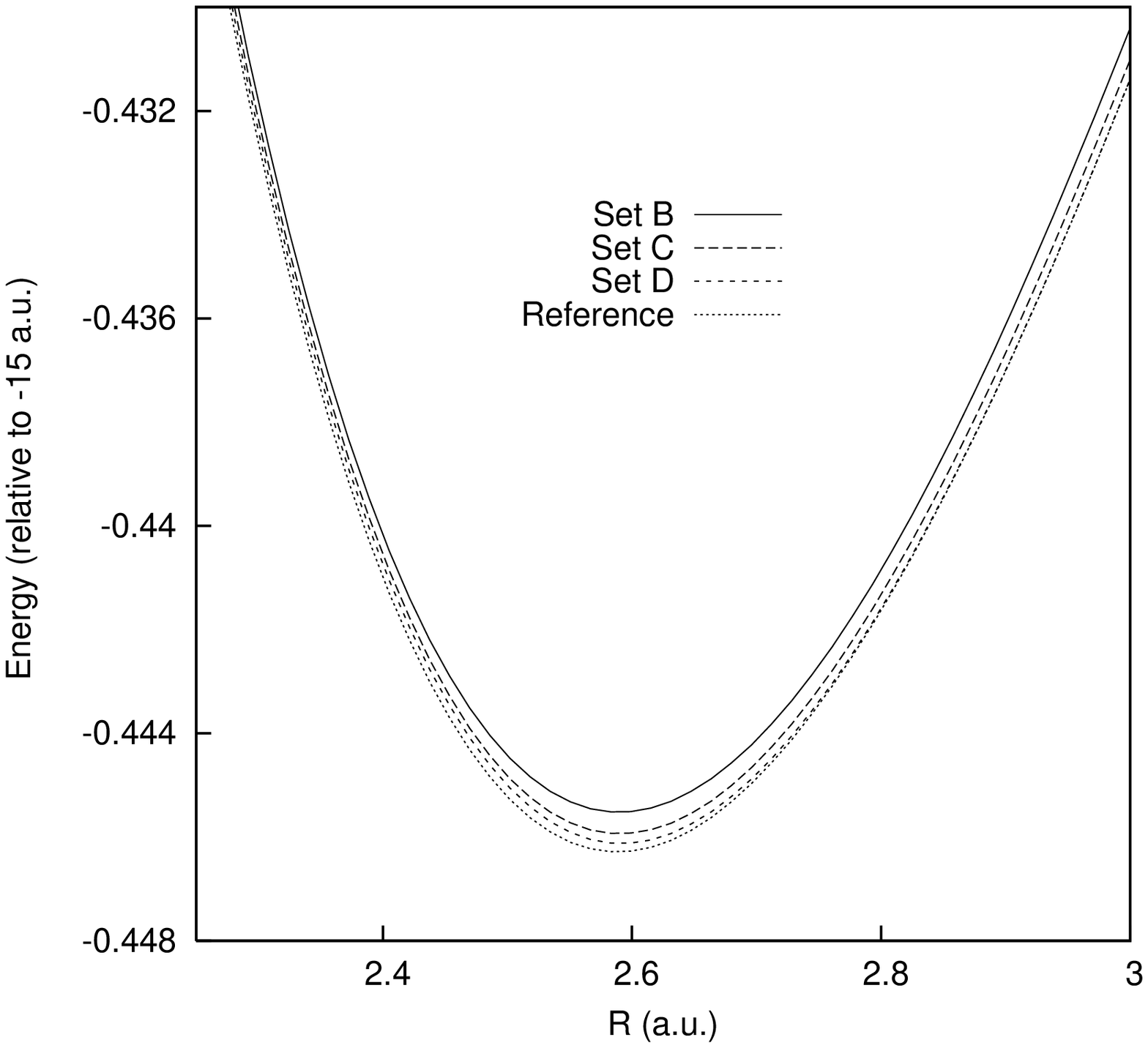}
\end{minipage}%
\caption{Potential energy curves for Cl$_2$(left panel) and HCl(right panel) in a.u. CCG and ACG results are compared for LDA XC 
potential.}
\end{figure}

After demonstrating the dependability and validity of this method, now in Table II, a representative set of 5 atoms and molecules
(ordered in ascending orders of N) are presented to assess its performance for a larger set of many-electron systems within the LDA
framework. All the quantities, as above were monitored. However, in order to save space, only kinetic, total and potential energy 
as well as N are given and compared. In this and all 
other following tables, experimental geometries are taken from computational chemistry database \cite{johnson06}. These
are all done in grid E, which has been found to be quite satisfactory for Cl$_2$ and HCl. For brevity, we quote only the grid-DFT 
results for reference, omitting ``grid-free" DFT, as we have seen earlier that the two generate results of very similar accuracy.
Once again for all of these, excellent agreement is observed; for more details, see \cite{roy08}.  

\begingroup
\squeezetable
\begin{table}
\caption {\label{tab:table2}Kinetic energy, $\langle T \rangle$, potential energy, $\langle V \rangle$, 
total energy, $E$, and $N$ for selected atoms and molecules (in a.u.) within LDA. CCG and ACG results are compared.} 
\begin{ruledtabular}
\begin{tabular}{lrrrrrrrr}
System     & \multicolumn{2}{c}{$\langle T \rangle$} & \multicolumn{2}{c}{$-\langle V \rangle$} & 
\multicolumn{2}{c}{$-\langle E \rangle$}   & \multicolumn{2}{c}{$N$}  \\
\cline{2-3}  \cline{4-5} \cline{6-7} \cline{8-9} 
        & CCG   & Ref.~\cite{schmidt93} &  CCG & Ref.~\cite{schmidt93} &  CCG & Ref.~\cite{schmidt93}  
        & CCG   & Ref.~\cite{schmidt93} \\ 
\hline
NaH            & 0.56931   & 0.56912 & 1.29712 & 1.29697 & 0.72781  & 0.72785  
               & 1.99999 & 2.00005            \\
As             & 2.07461 & 2.07354 & 8.10154 & 8.10047 & 6.02693 & 6.02693 
               & 5.00000 & 4.99999                    \\  
H$_2$S         & 4.90204  & 4.90197 & 16.10707 & 16.10698 & 11.20503  & 11.20501 
               & 8.00000 & 7.99989         \\ 
Br$_2$         & 8.55754  & 8.55716 & 34.74793 & 34.74755 & 29.19039  & 29.19039 
               & 14.00000 & 14.00003        \\  
MgCl$_2$       & 11.62114 & 11.62208 & 42.34513 & 42.34621 & 30.72399 & 30.72413 
               & 16.00004 & 15.99957    \\ 
\end{tabular}                                                                               
\end{ruledtabular}
\end{table}
\endgroup

After studying the LDA XC functionals, we now focus into the more important and useful so-called non-local functionals. Well-known
problems and discomfitures of LDA functionals for interacting many-electron systems are well documented in numerous communications
and it would be necessary to develop more accurate and elegant functionals for future application purposes. 
A frequently used and extremely successful candidate is the so-called BLYP 
\cite{becke88a,lee88} XC potential having dependence on gradients and Laplacian of density. This is a significant 
improvement over the LDA case and consequently has found many chemical, physical and biological applications. For practical 
implementation, however, it is preferable to use an equivalent form of the BLYP functional containing only first derivatives 
of density, as suggested
in \cite{miehlich89}. Following \cite{pople92}, this and other gradient-dependent functionals can be incorporated using a finite-orbital 
expansion method which helps avoiding the density Hessians. In the end, XC contribution of KS matrix is computed by the 
following expression,  
\begin{equation}
F_{\mu \nu}^{XC \alpha}= \int \left[ \frac{\partial f}{\partial \rho_{\alpha}} \chi_{\mu} \chi_{\nu} +
 \left( 2 \frac{\partial f}{\partial \gamma_{\alpha \alpha}} \nabla \rho_{\alpha} + 
    \frac{\partial f} {\partial \gamma_{\alpha \beta}} \nabla \rho_{\beta} \right) 
    \cdot \nabla (\chi_{\mu} \chi_{\nu}) \right] d\rvec,
\end{equation}
where $\gamma_{\alpha \alpha} = |\nabla \rho_{\alpha}|^2$, 
$\gamma_{\alpha \beta} = \nabla \rho_{\alpha} \cdot \nabla \rho_{\beta}$, $\gamma_{\beta \beta} = 
|\nabla \rho_{\beta}|^2$. This is advantageous because $f$ is a function \emph{only} of local quantities $\rho_{\alpha}$, 
$\rho_{\beta}$ and their gradients. All non-local functionals in this work are implemented using the Density Functional Repository 
program \cite{repository}. 

\begingroup
\squeezetable
\begin{table}      
\caption{\label{tab:table3} Comparison of BLYP energy components for Cl$_2$ and HCl in CCG and ACG, in a.u.} 
\begin{ruledtabular}
\begin{tabular} {lrrrrrr}
    & \multicolumn{3}{c}{Cl$_2$ ($R=4.2$ a.u.)}  &  \multicolumn{3}{c}{HCl ($R=2.4$ a.u.)}      \\
\cline{2-4} \cline{5-7} 
Set                        &     A      &    B       & Ref.~\cite{schmidt93}     &   A  &   B   &  Ref.~\cite{schmidt93} \\
$N_r$                      &    64      &   128      &             &   64        &    128      &             \\
$h_r$                      &    0.3     &   0.2      &             &   0.3       &    0.2      &             \\
$\langle T \rangle$        & 11.21504   & 11.21577   &  11.21570   & 6.25431     & 6.25464     & 6.25458     \\
$\langle V^{ne}_t \rangle$ & $-$83.72582& $-$83.72695& $-$83.72685 & $-$37.29933 & $-$37.29987 & $-$37.29979 \\
$\langle E_h \rangle$      & 36.74464   & 36.74501   &             & 15.86078    & 15.86103    &             \\
$\langle E_x \rangle$      & $-$5.29009 & $-$5.29015 &             & $-$3.01023  & $-$3.01026  &             \\
$\langle E_c \rangle$      & $-$0.37884 & $-$0.37892 &             & $-$0.21171  & $-$0.21174  &             \\
$\langle V^{ee}_t \rangle$ & 31.07572   & 31.07594   &  31.07594   & 12.63884    & 12.63903    & 12.63901    \\
$\langle E_{nu} \rangle$   &  11.66667  & 11.66667   &  11.66667   & 2.91667     & 2.91667     & 2.91667     \\
$\langle V \rangle$        & $-$40.98344& $-$40.98434& $-$40.98424 & $-$21.74382 & $-$21.74417 & $-$21.74411 \\
$\langle E_{el} \rangle$   & $-$41.43506& $-$41.43524& $-$41.43522 & $-$18.40618 & $-$18.40620 & $-$18.40620 \\
$\langle E \rangle$        & $-$29.76840& $-$29.76857& $-$29.76855\footnotemark[1] & 
                             $-$15.48951 & $-$15.48953 & $-$15.48953\footnotemark[2] \\
 $N$                       & 14.00006   & 14.00000   & 13.99998    & 8.00002     & 8.00000     &  8.00000    \\
\end{tabular}
\end{ruledtabular}
\footnotetext[1]{The \emph{grid-free} DFT value is $-$29.74755 a.u. \cite{schmidt93}.} 
\footnotetext[2]{The \emph{grid-free} DFT value is $-$15.48083 a.u. \cite{schmidt93}.} 
\end{table}
\endgroup

Table III presents a comparison of our CCG energy components for Cl$_2$ and HCl at $R=4.2$ and 2.4 a.u., respectively with BLYP
functional. A series
of calculations along the lines of Table I produced very similar conclusion we reached there for the LDA case. From these numerical
experiments, results for a few selected sets are given here, which is sufficient to illustrate the general trend. Again, reference
theoretical results correspond to those having same basis function, XC potential and effective core potentials. Both ``grid-DFT" 
results in ACG and ``grid-free" DFT results for total energy are reported for comparison. To convince us, some additional 
reference calculations are performed
for a decent number of atoms/molecules in various extended radial and angular grids besides the default grid of 
$N_r, N_{\theta}, N_{\phi} = 96,12,24$, namely, (i) $N_r, N_{\theta}, N_{\phi}=96,36,72$ (ii) $N_r,N_{\theta},N_{\phi}
=128,36,72$, with three integers denoting the number of integration points in $r,\theta,\phi$ directions respectively. For
the set of species chosen, all these grid sets offer results in very close agreement with each other. To summarize, 
out of 17 species, for 8 of them total energies remain unchanged up to 5th decimal place for all the grids. Very minor variations
were observed in remaining cases (largest deviation in total energy occurs for Na$_2$Cl$_2$, while for all others, it remains below
0.00007). However, as one passes from default grid to (ii), N gradually improves. As evident, CCG results are again in perfect
agreement with ACG results, as found for the LDA functional. Obviously Set B produces better results (absolute deviations being 0.00002
and 0.00000 a.u. for Cl$_2$ and HCl respectively) than Set A, but only marginally. Note that ``grid-free" and ``grid"-DFT results 
differ significantly from each other in this case and for all practical purposes, Set A suffices. Further details can be found in
\cite{roy08a}. 

As in the LDA case, next our calculated BLYP eigenvalues for Cl$_2$ and HCl (at same R values as in previous tables) are compared 
in Table IV. Clearly, all the eigenvalues for both molecules are very nicely reproduced, as observed for LDA functionals. 
Excepting the lone case of 3$\sigma$ levels (in both cases), where absolute deviation remains only 0.0001 a.u.,  Sets A,B 
results show a \emph{complete} matching for \emph{all} orbital energies. In the same vein of our LDA approach earlier, we also 
investigated total energies and other energy components for a broad range of internuclear distance of Cl$_2$ ($R=3.5-5.0$ a.u) and 
HCl ($R=1.5-3.0$ a.u.) both with 0.1 a.u. interval for BLYP XC functional. It is very satisfying that for both of them,
Sets A,B practically coincide with the reference values for the \emph{entire} range of $R$. For Cl$_2$, maximum 
absolute deviations of 0.0001 and 0.0002 a.u. have been recorded from Sets B,A respectively. For HCl, the same remains well below 
0.0001 a.u. only. A more thorough discussion is given in \cite{roy08a}.  

\begingroup
\squeezetable
\begin{table}
\caption{\label{tab:table4} Comparison of BLYP negative eigenvalues (in a.u.) of Cl$_2$, HCl with 
reference values at $R=4.2$ and 2.4. CCG and ACG values are given.} 
\begin{ruledtabular}
\begin{tabular} {lccclccc}
    MO   & \multicolumn{3}{c}{Cl$_2$ ($R=4.2$ a.u.)} & MO  & \multicolumn{3}{c}{HCl ($R=2.4$ a.u.)} \\
\cline{2-4}  \cline{6-8}
Set          & A      &   B     & Ref. \cite{schmidt93} &  &  A       & B       & Ref. \cite{schmidt93} \\ 
 $2\sigma_g$ & 0.8143 & 0.8143  & 0.8143   & $2\sigma$     &  0.7707  & 0.7707  & 0.7707   \\
 $2\sigma_u$ & 0.7094 & 0.7094  & 0.7094   & $3\sigma$     &  0.4168  & 0.4167  & 0.4167   \\
 $3\sigma_g$ & 0.4170 & 0.4171  & 0.4171   & $1\pi_x$      &  0.2786  & 0.2786  & 0.2786   \\
 $1\pi_{xu}$ & 0.3405 & 0.3405  & 0.3405   & $1\pi_y$      &  0.2786  & 0.2786  & 0.2786   \\
 $1\pi_{yu}$ & 0.3405 & 0.3405  & 0.3405   &  &            &          &                    \\ 
 $1\pi_{xg}$ & 0.2778 & 0.2778  & 0.2778   &  &            &          &                    \\
 $1\pi_{yg}$ & 0.2778 & 0.2778  & 0.2778   &  &            &          &                     \\
\end{tabular}
\end{ruledtabular}
\end{table}
\endgroup

Now Table V compares various energy components (only the kinetic, potential and total energies) for a few selected atoms and 
molecules with BLYP XC functional (ordered in increasing N as we descend the table). A smaller grid of $N_r =64, h_r=0.4$ was 
sufficient for atoms, whereas a larger grid $N_r=128, h_r=0.3$ was used for molecules. Other energy components are omitted for
brevity, as they show very similar agreements with literature values as observed in previous occasions. The overall agreement for 
these with reference values is excellent. For a set of 5 atoms and 10 molecules, total energies are found to be identical to those
from reference values in 5 occasions; otherwise, the maximum absolute deviation remains below 0.0013\%. Results for more atoms
and molecules as well as further observations on this can be found in \cite{roy08a}. 

\begingroup
\squeezetable
\begin{table}
\caption {\label{tab:table5}BLYP energy components (in a.u.) and $N$ for selected atoms, 
molecules with reference results. CCG results are compared with ACG.} 
\begin{ruledtabular}
\begin{tabular}{lrrrrrrrr}
System     & \multicolumn{2}{c}{$\langle T \rangle$} & \multicolumn{2}{c}{$-\langle V \rangle$} & 
\multicolumn{2}{c}{$-\langle E \rangle$}   & \multicolumn{2}{c}{$N$}  \\
\cline{2-3}  \cline{4-5} \cline{6-7} \cline{8-9} 
        & CCG   & Ref.~\cite{schmidt93} &  CCG & Ref.~\cite{schmidt93} &  CCG & Ref.~\cite{schmidt93}  
        & CCG   & Ref.~\cite{schmidt93} \\ 
\hline 
Na$_2$        & 0.14723  & 0.14723  &  0.52871 & 0.52871  & 0.38148  & 0.38148  & 1.99999  & 2.00000  \\ 
P             & 2.38891  & 2.38890  &  8.78249 & 8.78248  & 6.39358  & 6.39358  & 4.99999  & 4.99999  \\     
NaCl          & 5.83959  & 5.83957  & 21.01698 & 21.01694 & 15.17739 & 15.17737 & 8.00003  & 7.99999  \\
PH$_3$        & 4.18229  & 4.18224  & 12.40101 & 12.40096 & 8.21871  & 8.21872  & 7.99999  & 7.99999  \\  
H$_2$S$_2$    & 8.88238  & 8.88240  & 30.16535 & 30.16538 & 21.28297 & 21.28298 & 13.99999 & 13.99999 \\
\end{tabular}                                                                               
\end{ruledtabular}
\end{table}
\endgroup

Finally, Table VI presents the HOMO energies, $-\epsilon_{\mathrm{HOMO}}$ and atomization energies for selected 7 molecules 
in LDA and BLYP approximation, at their experimental geometries taken from \cite{johnson06}. From the above discussion, as
expected, reference theoretical results are practically identical to those obtained from current CCG work and thus omitted;
while available experimental values \cite{afeefy05}, wherever possible, are quoted appropriately. Experimental atomization 
energies with asterisks denote 298$^{\circ}$K values; otherwise they refer to 0$^{\circ}$K. Here ionization energies obtained from
a modified Leeuwen-Baerends (LB) exchange potential \cite{leeuwen94,schipper01}, with LDA correlation is also included 
for comparison. It may be noted that LDA and GGA XC potentials suffer from incorrect asymptotic long-range behavior; thus 
although ground-state total energies of atoms, molecules, solids are obtained quite satisfactorily, ionization energies and
higher-lying states are described rather poorly. The former is typically off by 30--50\% from experimental results. Note that  
our long-term objectives is to investigate the feasibility and applicability of this method for dynamical studies such as 
laser-atom/molecule interaction through such effects as multi-photon ionization, high-order harmonic generation, photo-ionization, 
photo-emission, photo-dissociation, etc., within a TDDFT framework. This is a very active, fascinating and challenging 
area of research from both experimental and theoretical point of view. These processes offer a host of important, fundamental
physical and chemical phenomena occurring in such systems and also they have found diverse practical applications (see, for 
example, \cite{brabec00,udem02,baltuska03,stapelfeldt03,posthumus04,gohle05}). For such studies, it
is necessary that both ionization energies and higher levels be approximated as accurately as possible, which is unfortunately 
not satisfied
by either LDA or BLYP functionals. The modified LB potential \cite{leeuwen94, schipper01}, $v_{xc\sigma}^{LB\alpha} 
(\alpha, \beta: \rvec)$, containing two empirical parameters, seems to be a very good choice in this case and as such, given by,
\begin{equation}
v_{xc\sigma}^{LB\alpha} (\alpha, \beta: \rvec) = \alpha v_{x\sigma}^{LDA} (\rvec) + v_{c\sigma}^{LDA} 
(\rvec) +
\frac{\beta x_{\sigma}^2 (\rvec) \rho_{\sigma}^{1/3} (\rvec)} 
{1+3\beta x_{\sigma}(\rvec) \mathrm{ln} \{x_{\sigma}(\rvec)+[x_{\sigma}^2 (\rvec)+1]^{1/2}\}}.
\end{equation}
Here $\sigma$ signifies up/down spins while the last term containing gradient correction is reminiscent of the exchange functional 
of \cite{becke88a}. $x_{\sigma}(\rvec) = |\nabla \rho_{\sigma}(\rvec)|[\rho_{\sigma}(\rvec)]^{-4/3}$ is a dimensionless quantity, 
$\alpha=1.19, \beta=0.01$. In this approximation of exchange, asymptotic long-range property is satisfied properly, 
i.e., $v_{xc\sigma}^{LB\alpha} 
(\rvec) \rightarrow -1/r, r \rightarrow \infty.$ Ionization energies obtained from LBVWN (LB exchange+VWN correlation), 
reported in Column 4, demonstrates its superiority over both LDA and BLYP functionals quite convincingly. LDA values are
consistently lower than the corresponding BLYP values while LBVWN energies are lower and more closer to experimental values
than both of these. Now calculated atomization energies in columns 6,7 show significant deviations from experimental results. 
In several occasions, surprisingly LDA atomization energies are better than their BLYP counterparts. However, this should not be used to
conclude the superiority of former over the latter, because there may be some cancellation of errors and also other factors
such as more accurate basis functions, core potentials, etc., should be taken in to consideration. Such deviations are not 
uncommon in DFT, though. Even very elaborate extended basis set all-electron calculations on several molecules show very 
large errors in a recent work \cite{cafiero06}. However this is an on-going activity and does not directly interfere with the
main objective of this work.             

\begingroup
\squeezetable
\begin{table}
\caption {\label{tab:table6} Comparison of negative HOMO energies, $-\epsilon_{\mathrm{HOMO}}$ (in a.u.) and atomization energies 
(kcals/mol) for selected molecules with LDA, LBVWN (LB+VWN) and BLYP XC functionals. Experimental results \cite{afeefy05} are also
given, wherever possible. An asterisk indicates 298$^{\circ}$K values. All others correspond to 0$^{\circ}$K.} 
\begin{ruledtabular}
\begin{tabular}{lccccccc}
Molecule & \multicolumn{4}{c}{$-\epsilon_{\mathrm{HOMO}}$ (a.u.)} & \multicolumn{3}{c}
{Atomization energy (kcals/mol)}  \\
\cline{2-5}  \cline{6-8}  
              &  LDA    & BLYP    &  LBVWN  & Expt.~\cite{afeefy05}    &  LDA     &  BLYP    &    
Expt.~\cite{afeefy05}   \\
\hline 
NaBr          & 0.1818  & 0.1729  & 0.3057  & 0.3050  & 87.47    &  78.94   &    86.8*   \\ 
SiH$_4$       & 0.3188  & 0.3156  & 0.4624  & 0.4042  & 339.43   &  312.02  &    302.6   \\
S$_2$         & 0.2007  & 0.2023  & 0.3443  & 0.3438  & 56.75    &  52.47   &    100.8   \\
BrCl          & 0.2623  & 0.2537  & 0.4133  & 0.4079  & 44.95    &  25.41   &    51.5    \\
AlCl$_3$      & 0.3081  & 0.2976  & 0.4603  & 0.4414  & 278.02   & 232.88   &    303.4   \\
P$_4$         & 0.2712  & 0.2575  & 0.3964  & 0.3432  & 200.77   & 142.99   &    285.9   \\
PCl$_5$       & 0.2825  & 0.2722  & 0.4397  & 0.3748  & 246.22   & 145.33   &    303.2   \\
\end{tabular}                                                                              
\end{ruledtabular}
\end{table}
\endgroup

\section{Concluding remarks}
We presented an alternate route for atomic/molecular calculation using CCG, within the framework of GTO-based LCAO-MO 
approach to DFT. Although several attempts in real-space are known which use CCG, however, to my knowledge, this is the 
first time such studies are made in a basis-set approach, \emph{solely} in CCG. Accuracy and reliability of our 
method is illustrated for a cross-section of atoms/molecules through a number of quantities such as energy components, 
potential energy curve, atomization energy, ionization potential, eigenvalue, etc. For a large number of species, these 
results virtually coincide with those obtained from other grid-based or grid-free DFT methods available. The success of this approach
lies in 
an accurate and efficient treatment of the Hartree potential, computed by a Fourier convolution technique 
by partitioning the interaction in to long-range and short-range components. No auxiliary basis set is invoked in to the picture. 
Detailed comparisons have been made which shows that the present results are variationally bounded.

\section{acknowledgments}
I thank professors D.~Neuhauser, S.~I.~Chu, E.~Proynov for useful discussions.

\bibliography{refn.bib}
\bibliographystyle{unsrt}
\end{document}